# On Removing the Classical-Quantum Boundary

Khaled Mnaymneh[1]

[1]National Research Council Canada, 1200 Montreal Road, Ottawa, Ontario Canada, K1A 0R6

**Abstract**: We argue that it is the assumption of counterfactual definiteness and not locality or realism that results in Bell inequality violations. Furthermore, this assumption of counterfactual definiteness is not supported in classical mechanics. This means that the Bell inequality must fail classically, effectively removing the classical-quantum boundary – a conclusion prophesized by Bell himself. An implication here is that a *local* hidden variable theory, in the configuration space of classical mechanics cannot be ruled out. One very surprising result is that classical mechanics, in the context of Hamilton's stationary principle, may in fact have stronger correlations than quantum mechanics, in that it may be the key to beat Tsirelson's bound.

**Introduction**: Hamilton's principle of stationary action states that from all possible paths in configuration space, Nature selects the one that is an extremum [1,2]. The condition for such a path is independent of the choice of coordinate system [3]. To transform to configuration space, we take the configuration of an *n*-particle system at some moment in time and represent it by a single particle, known as the *C*-point, within a 3*n*-dimensional space. Hamilton's principle is only satisfied when the generalized coordinates obey the Euler-Lagrange equations [4]. When we transform these equations back to our physical three-dimensional space, we have the familiar vector equations of Newton's second law [5].

While the metaphysics of Hamilton's principle [6] is beyond the scope of this work, its coordinate-independent nature establishes a type of dynamical indistinguishability and equivalence between configuration space and physical three-dimensional space. In other words, stationary action implies a high degree of correlation between all the physical dynamics occurring in our world. Hence, we posit that any definition of reality should include it. However, in 1935, when Einstein, Podolsky and Rosen set out to demonstrate that quantum mechanics was incomplete, they did not include it [7]. They then preceded to build a gedankenexperiment under their definition of reality and demonstrated a paradox proving quantum mechanics was incomplete. Nearly thirty years later, John Stewart Bell published a proof [8] demonstrating that any local hidden variable theory in physical three-dimensional space used to complete quantum mechanics [9] cannot be Lorentz invariant; there must be non-casual influences on distant detectors. Subsequent experiments [10, 11, 12] in the years that followed supported this proof by repeatedly demonstrating loophole-free [13] violations of his famous inequality. This established an experimental basis for what Schrödinger coined quantum entanglement [14]; the phenomenon that now underpins the promise of future technologies aimed at next-generation applications in quantum information processing and quantum encryption [15].

Recently, however, there have been experimental reports of *classical* entanglement [16] that has also triggered debate [17] in the literature. The nature of these arguments is centered on where the boundary between quantum and classical mechanics is located [18]. The experimental violations of Bell's inequality were regarded as the experimental foundation of this boundary. In this Letter, we argue, however, that this boundary does not exist nor has it ever existed. By including Hamilton's principle in the definition of reality Bell's inequality must also fail classically. Already in classical mechanics, motion is described as occurring on as symplectic manifold [3], where the term symplectic has an etymology [19] that implies *braided or intertwined*. Conditions for stationary action "braid" reciprocally-related variables into conjugate variables that obey the Poisson brackets, a non-zero area in phase space that must exist if the motion is deemed canonical (i.e., obey Hamilton's principle).

The EPR paper proves quantum mechanics to be incomplete, albeit counterfactually. If reality strictly lacks counterfactual definiteness (CFD), we hold that *quantum entanglement* can then be regarded as *classical intertwinement*; Bell's inequality must not hold in configuration space even though locality and realism hold there by definition. There is no other particle to influence the *C*-point, locally or non-locally. And all acts of measurement are represented by the *C*-point. A no-go theorem for local hidden variable theory does then not apply invariantly across all spaces. It would be impossible to produce a proof only because the coordinate-independent extremum makes locality, and the lack thereof, artifacts of the spaces one chooses to represent reality. However, if the lack of CFD is true in all spaces, Bell's inequality fails classically as well.

**Body**: There is debate in the literature as to whether or not Bell made a direct assumption of CFD [20,21]. The majority opinion, however, is that locality and realism are the offending assumptions, and that the lack of CFD is derived and not outrightly presumed [22]. A recent experiment [23] explored the assumption that some non-local influences could act during correlation measurements. The associated inequality [24] was shown to be experimentally violated. One of the possible explanations is that our world may lack CFD. We show that there is a foundational basis in classical mechanics to support this possibility. If Bell's inequality is violated in both classical and quantum mechanics, then their boundary ceases to exist. Following John Stewart Bell,

> "*More plausible to me is that we will find that there is no boundary. It is hard for me to envisage intelligible discourse about a world with no classical part – no base of given events, be they only mental events in a single consciousness, to be correlated. On the other hand, it is easy to imagine that the classical domain could be extended to cover the whole.*" [25]

*We seek here then to extend the classical domain to cover the whole.* One way for our world to strictly lack CFD is the nonexistence of initial conditions [26]. Consider a section of a possible actual path in configuration space between some point $q$ at some time $t$ and another, different, point $q_0$ at an earlier time $t_0$ [5]. The action along this section is defined as the time integral of the Lagrangian,

$$S[q(t)] = \int_{t_0}^{t_1} dt\, L(q(t), \dot{q}(t), t), \qquad (1)$$

where $S$ is the action, $q(t)$ and $\dot{q}(t)$ are the $f$ generalized position and velocity at some time $t$ in configuration space, and $L$ is the Lagrangian occurring in the interval $dt$ between moments in time. Assuming that we have completely solved the Euler-Lagrange equations, the generalized position function satisfying Hamilton's principle,

$$q(t) = f(t, c), \qquad (2)$$

where the $c$'s are a set of $2f$ integration constants. These two points in configuration space are correlated via the Hamilton's principal function (HPF) $S_H$,

$$S_H(q, t; q_0, t_0). \qquad (3)$$

The HPF is also known as type-1 generating function [5]. The conjugate momenta of this correlation can be found to be,

$$p_0 = -\frac{\partial S_H}{\partial q_0}, \quad p = \frac{\partial S_H}{\partial q}, \qquad (4)$$

where $p_0$ is the momentum C-point leaves position $q_0$ at time $t_0$ in order to reach position $q$ at time $t$ with momentum $p$. *Note that the position and momentum of the C-point are both definite*. These two moments in time have two associated "Hamilton-Jacobi"-like partial differential questions,

$$H_0 - \frac{\partial S_H}{\partial t_0} = 0, \quad H + \frac{\partial S_H}{\partial t} = 0, \qquad (5)$$

where $H_0$ and $H$ are the classical time-dependent Hamiltonians associated with those times.

The existence of these two equations having to be solved simultaneously poses an unsolved and serious problem [27]. Either endpoint can be made to be the initial value of the other whilst both endpoints sit on a one-dimensional infinity, a line, of initial values points all of which lead to the other endpoint [5]. We see that the laws of physics do not supply an ability to solve for the HPF directly. Jacobi [28] proposed that only one equation needs to be solved if the other phase space point is a constant in time. This yields the famous Hamilton-Jacobi equation,

$$H(q,p,t) + \frac{\partial}{\partial t} S_J(q;\alpha;t) = 0, \tag{6}$$

where $S_J$ is the type-2 generating function from position $\beta$, having momentum $\alpha$, to position $q$ having momentum $p$ at time $t$. The HPF could then be built from solving equation (6) for different constant points:

$$S_H(q,t;q_0,t_0) = \sum_i S_J(q;\alpha_i;t) - S_J(q_0;\alpha_i;t_0), \tag{7}$$

where the first $S_J$ is the type-2 function from constant position $\beta_i$ to position $q$ at time $t$, and the second $S_J$ is the type-2 function from constant position $\beta_i$ to position $q_0$ at time $t_0$. It is important to note that equation (7) breaks down when the Hamiltonians [5] are assumed to be time independent.

It is interesting to note that when Schrödinger wanted to make the Hamilton-Jacobi equation into an eigenvalue problem [29], he proposed the following form of Jacobi's generating function,

$$S_J(q;\alpha;t) \equiv K \log \psi(q,t), \tag{8}$$

where $K$ is the constant of proportionality, and $\psi(q,t)$ is a function that makes the Hamilton-Jacobi equation separable. Equation (8) looks similar to Boltzmann's statistical definition of entropy based on the number of microstates that the macrostate could possibly be arranged from.

We infer here that such a prescription [30] implies the impossibility of access to information. We posit that this lack of access is foundational because *there are no initial conditions to our reality*. On its own, this statement is extremely provisional, and contrary to currently accepted cosmological theories [31], even though there are new models that support alternative theories that lack a beginning [32,33]. However, our assertion is also strongly motivated by the experimental violations of Bell's inequality and the supposition that our world lacks CFD.

To connect the lack of CFD with Bell's inequality, we review how the inequality was produced. Using conservation rules and the space quantization of quantum spins, Bell proceeded to set up a bound on the expected difference in counterfactual joint deflections,

$$|E(\mathbf{a},\mathbf{b}) - E(\mathbf{a},\mathbf{c})| \leq 1 + E(\mathbf{b},\mathbf{c}), \tag{9}$$

where $E$ is the expected joint Stern-Gerlach deflections given the settings, $\mathbf{a}$, $\mathbf{b}$ and $\mathbf{c}$ of detectors. Note that $\mathbf{c}$ is a distant counterfactual setting [8]. Now, the assumption of CFD is clearly intended by Bell [34] when he writes,

> "We are not at all concerned with sequences of measurements on a given particle, or of pairs of measurements on a given pair of particles. We are concerned with experiments in which for each pair the 'spin' of each particle is measured **only once**." [emphasis by the author]

The implication here is that all measurement happens once, *including distant counterfactual settings*. Equation (9) can then be understood to mean that unperformed results bound performed results. However, if CFD is strictly lacking in our world, then unperformed experiments do not have results [35]. It is this bound that is violated in physical three-dimensional space and gives rise to the phenomena of entanglement. However, as we have proposed above, without initial variables, counterfactual definiteness is absent in classical configuration space and so this bound is also *invalid classically*.

Equation (9) is shown in the usual way to be violated for certain detector settings when the expectation values predicated in quantum mechanics are inserted within the inequality. However, quantum mechanics ultimately deals with wavefunctions that were conceived in configuration space. This is reasonable because when we use quantities defined in one space (i.e., wavefunctions conceived in configuration space), in an expression built in another space (i.e. our physical three-dimensional space), without including the appropriate transformations, we cannot reasonably expect the expression to hold for all values of variables.

We normally do not see a failure; and when we do, it is deemed "quantum". However, when the inequality is not violated, we are explicitly not including the rest of the universe in the "classical" result, we do not include the "whole" $C$-point. To do so, the configuration of the distant detector settings must be considered. When included, the classical expression equals the quantum expectation [8]. In fact, the system dynamics we are following will not be stationary unless the rest of the dynamics of the universe is added to it,

$$\delta S = \delta \left( S_{sys} + S_{uni} \right) = 0, \tag{10}$$

where $S_{sys}$ is the action of the system, $S_{uni}$ is the action of the rest of the universe and $S$ is the action of the sum. When we do see the inequality violated, and it appears to not include the rest of the universe, we must remember that in these experiments, we physically endeavor to carefully arrange the environment so that the entanglement phenomena can become apparent, whether by conducting experiments near absolute zero or using post-selection, etc. The hallmark attribute of entanglement is the irrelevance of space between the entangled parties, appearing as if they existed on the same zero-dimensional point; the $C$-point. By working to exclude the the environment from the system understudy, that particular modality of the experiment includes the environment coherently [36] so that phenomena like entanglement and teleportation can de demonstrated [37].

Endeavoring to consider how measurement modality and environment must be involved in the definition of reality can be traced back to Bohr's response [38] to the EPR paper [7]. He notes that instrument modality is fundamentally convolved with any measurement results, and that this convolution cannot be deconvolved in anyway. It enters into the very definition of the result, making any judgement on counterfactual experimental results irrelevant and unphysical. The inability to deconvolve is from the fact that the measurement and the subject to be measured belong to the same $C$-point in configuration space. The area of quantum contextuality [39,40], the result that tells us that measurement results depend on *how* the subject is measured, is directly related to Bohr's point. If we then add that the extremum path lacks a beginning, then any discussion about counterfactual measurements is unphysical and cannot be part of the definition of reality [7].

**Discussion**: The violation of Bell's inequality, and by how much it is violated up to the Tsirelson's bound [41], is now a measure the *quantumness* of some system, and a no-go theorem for a local hidden variable theory. However, we have shown that the assumptions of locality and realism are not necessarily questionable if that space is the configuration space of classical mechanics. In fact, when there is a single particle, there is no difficulty in giving a hidden variable account [8]. What is suspect in all spaces is the assumption of CFD. It was implicitly assumed in the EPR paper [7] and Bell directly assumed it [25] in the creation of his bound [8]. We have also shown that CFD is not possible in classical mechanics, if the extremum path runs from eternity to eternity. Even though it is impractical to prove this, CFD being the offending assumption in both classical and quantum mechanics could corroborate such a situation; if there is no way to "get started" then there can be no options in *how* to "get started".

If Bell's inequality is violated because of the lack of CFD in classical mechanics then question of a local hidden variable theory may still be an open question [42,43,44]. Bell [43], Kochen and Specker [39] did demonstrate that a local hidden variable theory in physical three-dimensional space was not possible; however, this is not to say that randomness dominates in this situation. In fact, quite the opposite: for 3/2-spin particles and greater, there can be no stochastic explanation for the associated spin eigenvalue equations [22]. One interpretation may be that the local hidden variable theory moves from three-dimensional space to configuration space, which may be where the subquantum-mechanical level of Bohm may be located [45]. It should be noted as well that elements of configuration space can be detected in physical three-dimensional space because the quantum wave function, which was conceived in configuration space [46], was detected in physical three-dimensional space [47].

Quantum technologies that promise solutions built upon entanglement need to consider incorporating quantum foundations in a rigorous way when promising solutions based on quantum mechanics. It must be understood that Schrodinger's example of a cat in box [48] was not to showcase the "strangeness of quantum mechanics" but his reason how interpretations of quantum mechanics can be fundamentally absurd. That said, the efforts and investments that has gone into quantum technology are neither lost nor wasted [49].

As a final point, this paper helps to identify potential launching points from where to further study stronger-than-quantum correlations [50] that seem to potentially put today's quantum technologies in jeopardy [51]. The Hamilton-Jacobi equation, the place where Schrödinger started, is itself an approximation to Hamilton's principle function. Could there be other lines of thought that can start where Jacobi did, instead of where Schrödinger stated? Perhaps, in a very unexpected and shocking way, classical mechanics, if we include relativity under classical mechanics, may hold the potential to outperform quantum mechanics; that classical intertwinement may be more correlated than quantum entanglement [50].

**Conclusion**: When the physical dynamics of a system is considered in configuration space, we see that the boundary between classical mechanics and quantum mechanics disappears. Counterfactual definiteness, or the lack thereof, does not arise due to the violation of locality and realism in configuration space but rather arises from the nonexistence of initial variables. We show this by reinterpreting Bell's inequality as proof against CFD, and not necessary against a local hidden variable theory. The nonexistence of the classical-quantum boundary where the classical domain is extended to the whole has profound implications for quantum technology. Bell showed that there no mathematical guarantee that hidden variables cannot exist. The existence of hidden variables, whether they are non-local in physical three-dimensional space or local in configuration space, expose a vulnerability that may still be exploited. Foundations, especially quantum foundations, need to be a vital part of any science program that promises to disrupt our technology in a profound and deep way.